# A Synonym Based Approach of Data Mining in Search Engine Optimization

Palvi Arora[1], Tarun Bhalla[2]
[1,2]Assistant Professor
[1,2]Anand College of Engineering & Management, Kapurthala

**Abstract:**

In today's era with the rapid growth of information on the web, makes users turn to search engines as a replacement of traditional media. This makes sorting of particular information through billions of webpages and displaying the relevant data makes the task tough for the search engine. Remedy for this is SEO (Search Engine Optimization), i.e. having a website optimized in such a way that it will display the relevant webpages based on ranking. This is the main reason that makes search engine optimization a prominent position in online world. This paper present a synonym based data mining approach for SEO that makes the task of improving the ranking of the website much easier way and user will get answer to their query easily through any of search engine available in market.

## 1. INTRODUCTION

### 1.1 Background on Data Mining

In simple terms, the process of transforming data into useful information is known as data mining. In other words it is mining of knowledge from data. A large amount of data is available these days due to increasing computerization and digitalization related to all areas like business, industry, science, finance, banking, healthcare, etc. Data mining is about finding insights which are statistically reliable, unknown previously, and actionable from data. **Data mining** (the analysis step of the "Knowledge Discovery in Databases" process, or KDD),[1] an interdisciplinary subfield of computer science,[2][3][4] is the computational process of Discovering patterns in large data sets involving methods at the intersection of artificial intelligence, machine learning, statistics, and database systems.[2] The overall goal of the data mining process is to extract information from a data set and transform it into an understandable structure for further use.[2] Aside from the raw analysis step, it involves database and database management aspects, data preprocessing, model and inference considerations, interestingness metrics, complexity considerations, post-processing of discovered structures, visualization, and online updating.

Data mining is performed with following tasks:

- **Outlier Analysis:** It is the identification of records which does not match the usual patterns. They might be interesting that require further investigation. They can assist in fraud detection in credit card transactions in banking sector.

- **Association rule learning**: It searches for relationships between various attributes. For example a computer sale shop may discover that printers are frequently bought along with desktops. So providing a good discount on combination of both can enhance the sales. This technique is quite useful for customer and market analysis.

- **Clustering**: It is the process of grouping together values in the data that have similar patterns, but these patterns are not known in advance. For example, in a bank customer data, after analyzing the data we may form clusters of customers who make more than ten transactions per week and other who make less than 10 transactions.

- **Classification**: It is the process of grouping the data into different classes on the basis of previously known structures. For example, a bank customer data may be classified according to region.

- **Regression**: Attempts to find a function which models the data with the least error. It fits the data onto the function so that one value can be derived from another.

- **Summarization**: It providing a more compact representation of the data set, including visualization and report generation.

### 1.2 Applications of Data Mining

The data mining includes applications in following areas:

- Medical and health care.
- Banking and finance
- Retail/Marketing
- Fraud detection
- Customer Management
- Search engine optimization Telecom industry
- Computer Security
- Education
- And many more

### 2. SEARCH ENGINES

Computer program that search database and internet sites for the documents containing keywords specified by a user. It is a website whose primary function is providing a search engine for gathering and reporting information available on the internet or a portion of the internet. Most popular search engines include Google, Bing, Yahoo, AOL search, Ask, Web crawler, Dog pile, Lycos Alta Vista, etc.

### 2.1 Goal of SEO

The purpose of SEO is to:
- Create a great, effective user experience.
- Communicate to the search engines your requirements so that they can recommend you most relevant websites related to your search.

### 2.2 Characteristics of Search Engine

- **Content:** Is searched by the requirement that is being given, the text on the page, and the titles and descriptions that are given.

- **Performance:** How fast is your engine and how fast it gives the results?
- **User Friendly:** How much is the engine user friendly. Is it easy to work with the interface? Does it have a high success rate?
- **Basic Search options:** Does it perform automatic default of AND assumed between words?
- **Results Display:** Are the results ranked by popularity or relevancy or both? A good search engine displays most relevant results on the top.
- **Consistency:** How consistent is it? Do you get different results at different times?

### 2.3 Working of Search Engine

When we use the term search engine in relation to the Internet, they are usually referring to the actual search forms that search through databases of HTML documents available all over the internet.

There are basically three types of search engines: Those that are powered by crawlers; ants or spiders and those that are powered by humans; and those that are a combination of the two.

Crawler-based search engines are those that use automated software (called crawlers) that visit a Web site, read the information on the actual website, read the site's meta tags and also follow the links that the site links to performing indexing on all linked Web sites as well. The crawler returns back all that information to a central depository, where the data is stored and indexed. The crawler will periodically return to the sites to check for any information that has updated. The frequency with which this happens is determined by the administrators of the search engine and it also affects the efficiency of the search engine.

Human-powered search engines depend on humans to submit information that is subsequently indexed and catalogued. Only information that is submitted is put into the index.

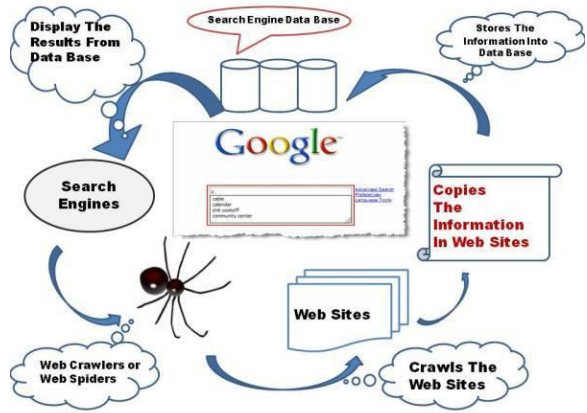

*Figure 1: Working of Search Engine*

### 3. SEARCH ENGINE OPTIMIZATION

Search Engine Optimization is the procedure of improving the visibility or traffic of a website or webpage in search engine via the natural [6] or unpaid searched results by increasing SERP (Search Engine Results Page) ranking. Optimization may target different types of search like image search, local search, video search, academic search, new search, industry specific vertical search .It can also be define as the process of affecting the visibility of a website or webpage in search engine.[8]

### 3.1 Importance of Data Mining in SEO

Internet is an immense, huge and dynamic data collection that includes infinite hyperlinks and volumes of data usage information-hence requires effective data mining. But huge data is still a challenge in knowledge discovery. [7]

• **Web pages are more complex than text data**: Web pages have dynamic data and do not follow any uniform structure. Web pages contains huge
amount of raw data that is not indexed therefore searching in web data has become more complex; time consuming and difficult.

• **The Web constitutes high quantity of dynamic information**: Web not only contains static data but also data that requires timely updating such as news, stock markets, live channels etc.

• **Web users include different kinds of user communities:** People from different communities have different backgrounds and use internet for different usage purposes. Many have different interests and lack knowledge of internet usage. Hence user gets lost within huge amount of data.

• **Only a small portion of the Web's pages contain truly relevant information**: A given user generally focuses on only a tiny portion of the Web, dismissing the rest as uninteresting data that serves only to swamp the desired search results.

### 3.2 Approaches for search engine

1. **Keyword-based search:** This includes search which use keyword indices or manually built directories to find documents with specified keywords or topics.e.g engines such as Google or Yahoo

2. **Querying deep Web sources:** Where information such as amazon.com's book data and realtor.com's real-estate data, hides behind searchable database query forms-that, unlike the surface web, cannot be accessed through static URL links.

3. **Random Surfing:** That follows web linkage pointers

### 4. PROPOSED METHOD

#### 4.1 Synonym Based Search
Today search is performed by searching the exact keywords entered by the user. But this may not result in the effective search because user may not know exact keywords. For example-search for data mining may not result in documents related to knowledge discovery, classification and outliers because these documents may not contain keyword data mining.

A search related to binary trees may not result in returning documents of AVL Trees and Red Black Trees.
Data mining can be effective in this methodology as it can help to classify synonyms related to a keyword.

#### 4.2 Architecture

A synonym table needs to be constructed which will map a query to the appropriate synonyms.
Hence it will enhance appropriate results. More appropriate synonyms will provide effective results because it will search for more documents containing the relevant data.

### 4.3 Synonym table

*Table 1: Synonym Table*

| Keyword | Synonyms |
|---|---|
| Car | Auto, vehicle |
| Data mining | Knowledge discovery |
| Airplane | jet |

A synonym table can be in the form of chained hash table in which a keyword entry will contain links to its synonyms.

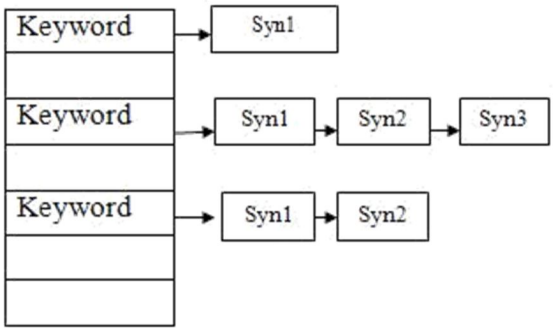

*Figure 2: Representation of Keywords with synonym*

### 4.4 Methodology

1. Query input
2. Association or identification of synonyms
3. Matching the keywords and synonyms on web
4. Displaying relevant results.

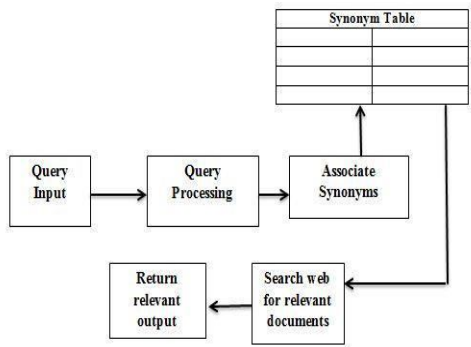

*Figure 3: Synonym Based Architecture*

1. **Query Input:** user inputs the query on which he wants to perform the search. User will mention keywords relevant to his query.

2. **Association of synonyms:** After the query is fed; it is then taken up for further processing. In this phase synonym table is looked up for the keyword entered and the corresponding
synonyms are fetched from chained hash table. Each entry of synonym will contain the pointer to the next synonym of the keyword; search will end when null pointer is encountered. Hence the list of synonyms relevant to keyword is retrieved.

3. **Matching the keywords and synonyms on web:** Web is searched for relevant documents based on keyword as well as synonym. Example- search for data mining will include search for data mining as well as knowledge discovery, classification, etc.

4. **Displaying relevant results:** Documents found relevant to query are returned and displayed on SERP (Search Engine Results Page).

### 4.5 Drawbacks

1. It may not be able to perform effective search for queries which have no relevant synonyms in the synonym table.
2. It is difficult to implement synonym table containing separate entry for every keyword because there is no limit to keywords that a user can search for.
3. Large amount of time can be consumed in looking up synonym table if a keyword has many synonyms.

### 5. CONCLUSION

In this paper we have proposed a search based on synonyms. This approach can be further extended or improved by implementing synonym table in a more effective way which should include less space consumption and minimum access time. This SEO approach can increase the ranking of a website on SERP and benefit its owners and provide users with more accurate and relevant search results.